\documentclass[epj]{svjour}

\usepackage[pdftex]{graphicx}
\usepackage{color}

\usepackage{amsmath}
\usepackage{amssymb}
\usepackage{braket}

\usepackage[sort&compress]{natbib}
\bibpunct{[}{]}{,}{n}{}{}

\newcommand{\fig}[1]{figure~\ref{fig:#1}}
\newcommand{\eq}[1]{(\ref{eq:#1})}
\newcommand{\lr}[1]{\ensuremath{\left( #1 \right)}}
\renewcommand{\d}{\mathrm{d}}

\renewcommand{\Im}[1]{\ensuremath{\mathrm{Im} \left(#1\right)}}
\newcommand{\I}{\mathrm{i}}
\newcommand{\ap}{\alpha}

\newcommand{\gm}{\gamma}
\newcommand{\Gm}{\Gamma}

\newcommand{\Dl}{\Delta}

\newcommand{\veps}{\varepsilon}
\newcommand{\sg}{\sigma}

\newcommand{\sech}{\mathrm{sech}}
\newcommand{\Abs}[1]{\ensuremath{\left| #1 \right|}}
\newcommand{\Tr}[1]{\ensuremath{\mathrm{Tr}\left(#1\right)}}

\begin{document}

\title{Localization under the effect of randomly distributed decoherence}
\author{Thomas Stegmann\inst{1}\fnmsep\thanks{\email{thomas.stegmann@uni-due.de}} \and Orsolya
  Ujs\'{a}ghy\inst{2} \and Dietrich E. Wolf\inst{1}}
\authorrunning{T. Stegmann et al.}
\institute{Department of Physics and CENIDE, University of Duisburg-Essen, D-47048 Duisburg, Germany \and
  Department of Theoretical Physics, Budapest University of Technology and Economics, H-1521 Budapest, Hungary}

\date{Date: 5 February, 2014\\}

\abstract{ Electron transport through disordered quasi one-dimensional quantum systems is studied. Decoherence
  is taken into account by a spatial distribution of virtual reservoirs, which represent local interactions of
  the conduction electrons with their environment. We show that the decoherence distribution has observable
  effects on the transport. If the decoherence reservoirs are distributed randomly without spatial
  correlations, a minimal degree of decoherence is necessary to obtain Ohmic conduction. Below this threshold
  the system is localized and thus, a decoherence driven metal-insulator transition is found. In contrast, for
  homogenously distributed decoherence, any finite degree of decoherence is sufficient to destroy
  localization. Thus, the presence or absence of localization in a disordered one-dimensional system may give
  important insight about how the electron phase is randomized.}

\maketitle

\section{Introduction} \label{sec:Intro}

The electron transport through nanosystems takes place in an intermediate regime between classical and quantum
transport \cite{Datta1997, Datta2005, Datta2012, Nazarov2009}. Thus, a quantum description of the system must
take the effect of decoherence into account. This raises then the fundamental, but up to now only partially
answered question, whether decoherence enhances or reduces transport. In this respect tight-binding chains
\cite{Thouless1981, Shi2001, Zilly2009, Zilly2012, Stegmann2012, Znidaric2013}, molecular wires
\cite{Zilly2010, Cattena2010, Nozaki2012} and aggregates \cite{Gaab2004, Plenio2008, Rebentrost2009,
  Kassal2012} were studied. Surprisingly it was found that the transport can be enhanced by decoherence, a
phenomenon referred to as \textit{decoherence-assisted transport}.  Recently, it has also been discussed, if
Anderson localization can be observed in the presence of many-body interactions \cite{Gornyi2005, Basko2006,
  Oganesyan2007, Znidaric2010_2}. As many-body interactions are a source of decoherence, this corresponds to
the question, if localization is possible under the effects of decoherence.

In this paper, we address this question using a statistical model for the effects of decoherence
\cite{Zilly2009, Zilly2010, Zilly2012, Stegmann2012}. We use \textit{virtual decoherence reservoirs}
\cite{Buettiker1986_1, Buettiker1991}, where the electrons are absorbed and reinjected after randomization of
phase and momentum. These decoherence reservoirs represent local interactions of the conduction electrons with
their environment. Their spatial distribution is governed by the underlying microscopic decoherence
processes. For example, when the decoherence is caused by random, uncorrelated scattering, also the
decoherence reservoirs are distributed in this way. Afterwards, the transport property of interest
(e.g. resistance or conductance) is ensemble averaged over the spatial distributions of decoherence
reservoirs. In contrast, when a continuous loss of the electron phase is considered, a homogenous distribution
of decoherence reservoirs is used. Continuous phase randomization was considered by Pastawski et
al. \cite{Amato1990, Cattena2010, Nozaki2012}. It is also used in other virtual reservoir approaches
\cite{Shi2001, Li2002, Yu2002, Maschke1994, Knittel1999, Pala2004, Zheng2006} and other phenomenological
models \cite{Gaab2004, Plenio2008, Rebentrost2009, Kassal2012, Golizadeh-Mojarad2007, Znidaric2010_2,
  Znidaric2013}. However, it is not clear if continuous decoherence is justified in every system.

\begin{figure}[t] 
  \centering
  \includegraphics[scale=1]{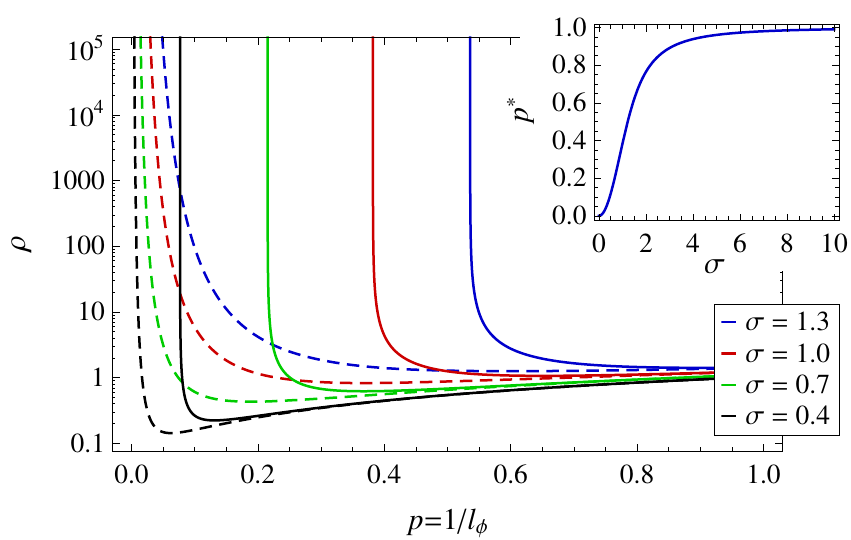}
  \caption{(Color online) Resistivity $\rho$ of an infinitely long disordered tight-binding chain as a
    function of the degree of decoherence $p$ (inverse phase coherence length $1/\ell_{\phi}$). If the
    decoherence is distributed homogenously (dashed curves), $\rho$ is Ohmic for any $p>0$. If the decoherence
    distribution is random and uncorrelated (solid curves), a minimal degree of decoherence $p^*$ is necessary
    to obtain Ohmic conduction, whereas below this threshold the system is localized ($\rho \to \infty$). $p^*$ as a
    function of the disorder $\sg$ is shown in the inset.}
  \label{fig:1}
\end{figure}

Studying disordered tight-binding chains and ribbons, here we explain why the distribution of the decoherence
reservoirs has a significant effect on the transport. When the decoherence reservoirs are distributed randomly
without spatial correlations, we find length-independent Ohmic resistivity only if the degree of decoherence
exceeds a certain threshold, or in other words, if the phase coherence length is sufficiently short. Otherwise
the system is localized, which is indicated by a divergence of the resistivity at a critical degree of
decoherence, see the solid curves in \fig{1}. When the decoherence reservoirs are distributed homogenously,
the transport is Ohmic for any finite degree of decoherence, see the dashed curves in \fig{1}. Our results may
help to learn from resistance measurements how the decoherence is distributed and how the electron phase
information is lost in the system.

\section{Quantum transport in presence of phase randomizing reservoirs} \label{sec:QuTrans}

In general, we consider electron transport through a quantum system, described by a single-particle
Hamiltonian $H$. The system is in contact with two real source and drain reservoirs as well as several
virtual decoherence reservoirs.

The current from the $j$th to the $i$th reservoir is calculated by means of the Landauer formula
\cite{Datta1997, Datta2005, Datta2012}
\begin{equation}
  \label{eq:1}
  I_{ij}= \frac{e}{h} \int \d E \, T_{ij} \lr{f_j - f_i},
\end{equation}
where $T_{ij}$ is the coherent transmission between the reservoirs and $f_i$ are the energy distribution
functions of the reservoirs. The energy-distribution functions of the source and drain reservoir are assumed
as Fermi functions $f_{S/D}$. The energy-distribution functions of the virtual reservoirs are determined by
the constraint that the total (energy-resolved) current at a virtual reservoir has to vanish. At an
infinitesimal bias voltage and at zero temperature, the total resistance of such a system, measured in units
of $h/e^2$, is given by (see e.g. \cite{Amato1990})
\begin{equation}
  \label{eq:2}
  R= \frac{1}{T_{DS} +\textstyle{\sum_{ij}} T_{Di} \mathcal{R}_{ij} T_{jS}},
\end{equation}
where
\begin{equation}
  \label{eq:3}
  \mathcal{R}^{\,-1}_{ij}=
  \begin{cases}
    -T_{ij} & i \neq j,\\
    \sum_{k \neq i} T_{ik} & i = j.
  \end{cases}
\end{equation}
The sum in \eq{2} is over the virtual reservoirs, whereas the sum in \eq{3} is over all reservoirs including
source and drain.

When virtual decoherence reservoirs are introduced within a one-dimensional quantum system and when complete
phase randomization is assumed, the coherent transmission $T_{ij}$ is limited to nearest neighbors and thus,
the system can be subdivided into smaller coherent subsystems. Simplifying \eq{2} under these conditions, the
resistance of the system is given by the sum of the subsystem resistances \cite{Zilly2009}
\begin{equation}
  \label{eq:4}
  R= \sum_{i} \frac{1}{T_{i+1,i}}.
\end{equation}

The above formulae for the resistance reflect Kirchhoff's law for networks of resistances $1/T_{ij}$, which
have to be calculated by quantum mechanics. Applying the non-equilibrium Green's function approach
\cite{Datta1997, Datta2005, Datta2012}, the transmission from the $j$th to the $i$th reservoir is given
by
\begin{equation}
  \label{eq:5}
  T_{ij}= 4\Tr{\Im{\Gm_i}G\Im{\Gm_j}G^+},
\end{equation}
where the Green's function is defined as
\begin{equation}
  \label{eq:6}
  G= \Big[E-H-\textstyle{\sum_k \Gm_k} \Big]^{-1}.
\end{equation}
The influence of each of the reservoirs is taken into account by a self-energy $\Gm_k$, which is in general a
complex-valued function and thus, connects the isolated quantum system to the environment.

Here, the decoherence is considered as a statistical process due to (dynamical) scattering. Thus, after
calculating the resistance by means of \eq{2} or \eq{4} for a given spatial distribution of decoherence
reservoirs, the transport property of interest (e.g. resistance or conductance) is ensemble averaged over the
decoherence distributions.

\section{Results} \label{sec:Res}

\subsection{Disordered tight-binding chains} \label{sec:TBCh}

We begin with tight-binding chains of length $N$, which are described by the Hamiltonian
\begin{equation}
  \label{eq:7}
  H= \sum_{i=1}^{N} \veps_i \ket{i}\bra{i} +\sum_{i=1}^{N-1} t\lr{\ket{i}\bra{i+1} + \ket{i+1}\bra{i}}.
\end{equation}
The coupling $t$ between neighboring sites is assumed to be homogenous and is used as the energy unit
$t=1$. The onsite energies $\veps_i$ are distributed independently according to a probability distribution
$w(\veps)$ with mean 0 and variance $\sg^2$. In order to keep the discussion clear and simple, we consider
here only the band-center $E=0$ and wide-band contacts $\Gm_k = -\I \ket{k}\bra{k}$, where $k$ denotes the
sites connected to the $k$th reservoir (i.e. an end of the chain). However, we stress that the main results of
this paper are still valid outside the band-center and for arbitrary self-energies.

As proved in the Appendix \ref{sec:DisAvRes} (including the general case of arbitrary energies $E$ and
self-energies $\Gm$), the disorder averaged resistance of the coherent chain of $N$ sites is given by the
compact analytical formula
\begin{align}
  \label{eq:8}
  \left\langle \frac{1}{T_N} \right\rangle&= \int \frac{1}{T_N} \prod_{i=1}^N w(\veps_i) \d \veps_i \notag\\
  &= \frac{1}{2}\lr{1 +\ap_+ e^{N/\xi} +\ap_- e^{-N/\xi} (-1)^N},
\end{align}
where $\langle \cdot \rangle$ denotes disorder averaging and
\begin{equation}
  \label{eq:9}
  2\ap_\pm= 1 \pm \sech{\lr{\xi^{-1}}},
\end{equation}
\begin{equation}
  \label{eq:10}
  \xi^{-1}= \log\lr{\frac{\sg^2}{2}+\sqrt{\frac{\sg^4}{4}+1}}.  
\end{equation}
As expected for a localized system, the disorder averaged resistance of the coherent chain increases
exponentially with its length. Contributions from a constant as well as from an exponentially suppressed
oscillatory term are irrelevant for $N \to \infty$. Note that in the limit $\sg \to 0$ we recover the length
independent resistance of a ballistic conductor $\braket{1/T_N}= 1$. 

\begin{figure}[t]
  \centering
  \includegraphics[scale=0.4]{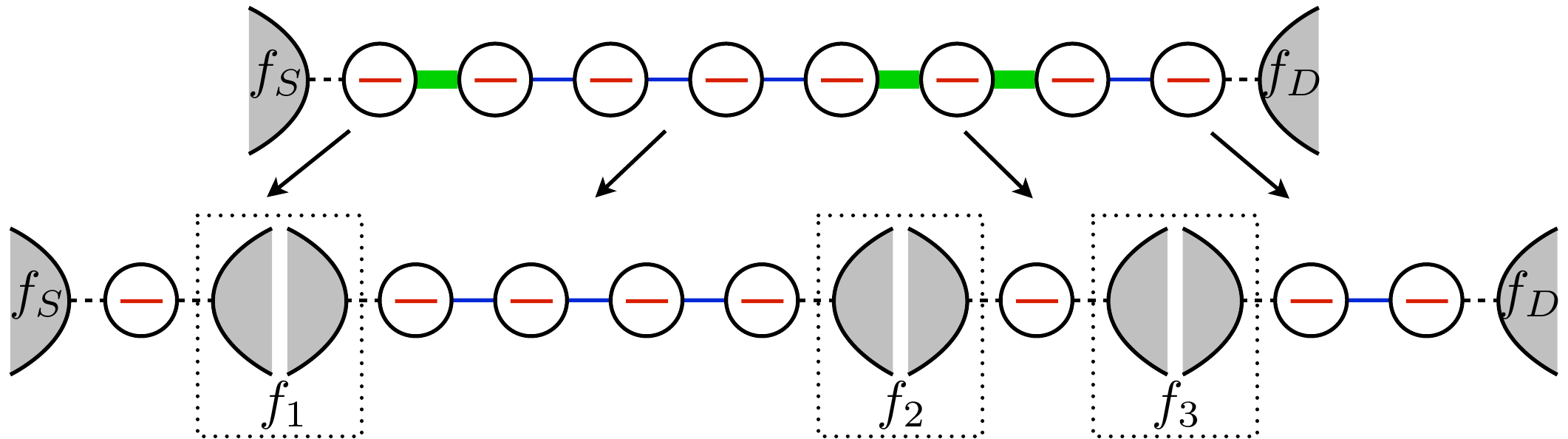}
  \caption{(Color online) Effects of decoherence are taken into account by replacing bonds of the
    tight-binding chain with virtual decoherence reservoirs, where phase and momentum are randomized
    completely. These assumptions allow to subdivide the system into smaller coherent subsystems.}
  \label{fig:2}
\end{figure}

For the effects of decoherence we replace bonds of the chain with completely phase and momentum randomizing
reservoirs\footnote{In Ref. \cite{Stegmann2012} we have shown that in a modified model pure dephasing can also
  be obtained, i.e. complete phase randomization but conservation of the momentum at a virtual
  reservoir. However, this would change the resistivity only by an additive constant and does not affect the
  main results of this paper.}, which subdivide the system into smaller coherent subsystems, see \fig{2}. We
focus on homogenous as well as on random uncorrelated distributions of the decoherence reservoirs, but allow
in general any arbitrary decoherence distribution under the constraint that on average $\lr{N-1}p$ decoherence
reservoirs are introduced in the chain. The \textit{degree of decoherence} $p$ can be related to the average
subsystem length and thereby defines the phase coherence length
\begin{equation}
  \label{eq:11}
  \ell_{\phi}= \frac{N}{1+\lr{N-1}p} \xrightarrow{N \to \infty} \frac{1}{p}.
\end{equation}

We start by ensemble averaging the resistivity of the chain over decoherence distributions. According to the
ergodic hypothesis, this corresponds to experiments where the current through the system is fixed and the
voltage drop is measured. Afterwards we will show that the main results of this paper are not changed, when the
conductivity is ensemble averaged over decoherence distributions.

For the random uncorrelated decoherence distribution, the bonds of the chain are replaced with probability $p$
by virtual reservoirs. The average number of coherent subsystems with length $j$ in a chain of length $N$ is
given by
\begin{equation}
  \label{eq:12}
  u_j= e^{-(j-1)/\ell}
  \begin{cases}
    2p+\lr{N-1-j}p^2 &\text{for } j<N,\\
    1 &\text{for } j=N.
  \end{cases}
\end{equation}
where
\begin{equation}
  \label{eq:13}
  \ell^{-1}= -\log\lr{1-p}.
\end{equation}
The resistivity of the chain
\begin{equation}
  \label{eq:14}
  \rho \equiv \frac{\left\{ \braket{ R }\right\}}{N}= \frac{1}{N}\sum_{j=1}^N u_j \left\langle \frac{1}{T_j}
  \right\rangle,
\end{equation}
ensemble averaged over decoherence $\{ \cdot\}$ and disorder $\langle \cdot \rangle$ configurations, is then
calculated by means of \eq{8} and \eq{12}. We can directly see from the products $u_j \langle 1/T_j\rangle$
that the resistivity is determined by the relation of the characteristic length $\xi$ and $\ell$, see the
Appendix \ref{sec:DphAvRes} for detailed derivations. When $\xi > \ell$ the transport is Ohmic, i.e. the
resistivity is length-independent
\begin{equation}
  \label{eq:15}
  \rho_{\xi>\ell} \xrightarrow{N \to \infty} p+ \frac{\sg^2}{4} \frac{p}{p-\sg^2\frac{1-p}{2-p}}.
\end{equation}
However, when $\xi<\ell$ the system is localized, i.e. its resistivity diverges exponentially with the chain
length
\begin{equation}
  \label{eq:16}
  \rho_{\xi<\ell} \propto  e^{\lr{1/\xi -1/\ell}N}.
\end{equation}
The root of the exponent $\xi^{-1}-\ell^{-1}=0$ determines the critical degree of decoherence
\begin{equation}
  \label{eq:17}
  p^* = 1 - e^{-1/\xi},
\end{equation}
where the transition between Ohmic and localized behavior appears. It can be related by \eq{11} to a critical
phase-coherence length and is a function of the disorder strength $\sg$, see \eq{10}. In a completely
different way, we obtained \eq{17} already in \cite{Zilly2012}. Here, its derivation by the new analytical
formula \eq{8} and \eq{12} allows to understand the statistical origin of the decoherence-induced
insulator-metal transition:

Localization is found to survive decoherence, when the exponentially increasing resistance of the long
coherent subsystems \eq{8} exceeds their exponentially decreasing frequency of occurrence \eq{12}. Any
decoherence distribution, for which the number $u_j$ of coherent subsystems decreases with their length $j$
faster than exponentially will show only Ohmic behavior. A simple example is to distribute the decoherence
reservoirs randomly under the constraint that at least after $j_{\text{max}}$ normal bonds a decoherence
reservoir has to be introduced. The corresponding $u_j$ has then a cut-off $u_{j>j_{\text{max}}}=0$ and thus,
decreases faster than exponentially. In this case the system is Ohmic for any finite degree of decoherence.
Also for a homogenous decoherence distribution, where all subsystems have the same size $\ell_\phi$, the
resistivity
\begin{equation}
  \label{eq:18}
  \rho_\text{hom}= \frac{1}{\ell_\phi} \left\langle \frac{1}{T_{\ell_\phi}} \right\rangle
\end{equation}
is Ohmic for any finite degree of decoherence.

If however, $u_j$ decreases with $j$ asymptotically more slowly than exponentially, the system will always be
localized, in spite of decoherence. This behavior appears for example, if the probability $p_j$ of having
coherent subsystems of length $j$ (i.e. $j-1$ succeding normal bonds) decreases as $p_j \propto j^{-\gm}$ with
an arbitrary constant $\gm>0$.

This strong influence of the decoherence distribution on the transport is the main result of this paper and is
summarized in \fig{1}, where the resistivity of the infinite long chain is shown as a function of the degree
of decoherence. The dashed curves for homogenous decoherence clearly show decoherence-assisted transport, and
agree qualitatively well with other studies assuming homogenous decoherence \cite{Thouless1981, Shi2001,
  Gaab2004, Plenio2008, Rebentrost2009, Kassal2012, Cattena2010, Nozaki2012, Znidaric2013}. The solid curves
for random uncorrelated decoherence exhibit divergencies at the critical degree of decoherence $p^*$, which is
shown as a function of the disorder strength $\sg$ in the inset of \fig{1}. Thus, in contrast to homogenous
decoherence, where the transport is Ohmic for any $p>0$, we find for random uncorrelated decoherence a
metal-insulator transition at $p^*$.

Taking many-body interactions explicitly into account, such a transition is found at a critical temperature
\cite{Gornyi2005, Basko2006}, which is proportional to the degree of decoherence in the system. Experiments on
various nanosystems have been performed, see e.g. \cite{Gershenson1997, Ahlskog1997, Khavin1998, Vavro2005,
  Ruess2007}, where a transition from Ohmic to exponential behavior is observed. It is found
\cite{Gershenson1997} that this transition occurs when the phase coherence length approaches the localization
length\footnote{Note that our parameter $\xi^{-1}$ is not the inverse localization length, which is generally
  defined as $\lim_{N \to \infty} \frac{1}{2N}\braket{\log T_N}$, but the second-order generalized Lyapunov
  exponent, defined as $\lim_{N \to \infty} \frac{1}{N}\log\braket{T_N}$, see \cite{Zilly2012}. Anyway, both
  quantities are a measure for the localization in the system.}, which agrees with our condition
\eq{17}. However, to our knowledge these experiments have been done only for systems with a fixed length. In
order to learn from an experiment how the decoherence is distributed, we propose to study the resistivity of
linear nanosystems as a function of their length. When the decoherence is randomly distributed, we expect that
below a critical temperature the resistivity increases exponentially with the chain length. Above the critical
temperature the resistivity should be constant (Ohm's law). In contrast, for homogenous decoherence we expect
for any non-zero temperature Ohmic behavior, when the length of the system is increased.

\subsection{Tight-binding ribbons and other model variations} \label{sec:TBRb}

In this Section we show that our results are not model specific, but appear more generally. We show that the
transition also appears, (I) when the phase is randomized only partially at a virtual reservoir, (II) when
tight-binding ribbons instead of chains are studied and, (III) when the conductivity is ensemble averaged
instead of the resistivity.

\begin{figure}[t] 
  \centering
  \includegraphics[scale=1]{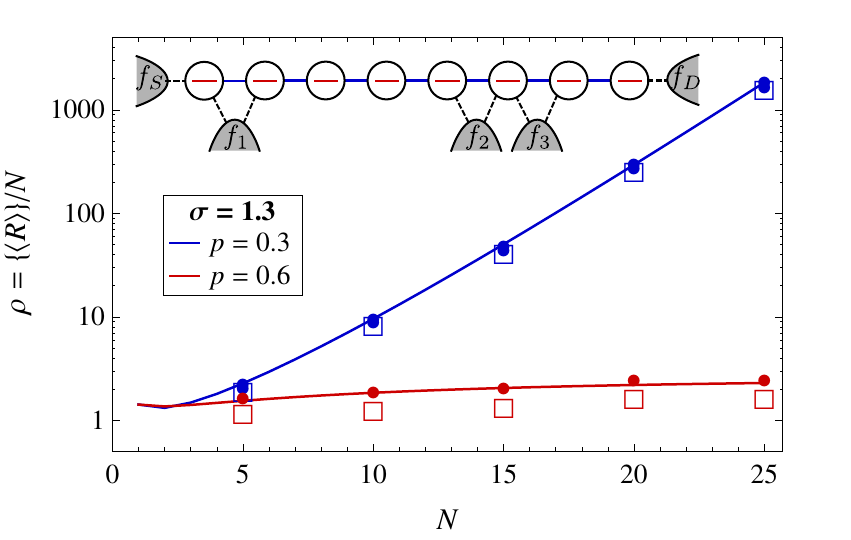}
  \caption{(Color online) Resistivity of the disordered chain as a function of its length. The decoherence
    induced transition also appears, when the randomly distributed decoherence reservoirs are only attached to
    the chain ($\square$). Averages were calculated numerically over $10^9$ decoherence and disorder
    configurations. Also shown is the average over the same ensemble under the assumption of complete phase
    randomization ($\bullet$), as well as the corresponding analytical result \eq{14} (solid curves).}
  \label{fig:3}
\end{figure}

When the assumption of complete phase randomization is abandoned by attaching the virtual reservoirs only to
the chain, see the inset of \fig{3}, the coherent transmission between all reservoirs has to be taken into
account and the ensemble average can be calculated only numerically by means of \eq{2} and \eq{3}.  However,
as shown in \fig{3}, the decoherence induced transition appears for attached reservoirs (squares) as well as
for completely phase randomizing reservoirs (circles). Contributions from the next-nearest neighbors can be
observed only for higher degrees of decoherence, as the coherent transmission between two reservoirs is
exponentially suppressed with their distance. The numerical averages agree well with the analytical result
\eq{14} for completely phase randomizing reservoirs, see the solid curves. Combining this result with \eq{18},
we can conclude that attaching a homogenous distribution of decoherence reservoirs to the chain leads to Ohmic
transport for any finite degree of decoherence.

\begin{figure}[t] 
  \centering
  \includegraphics[scale=1]{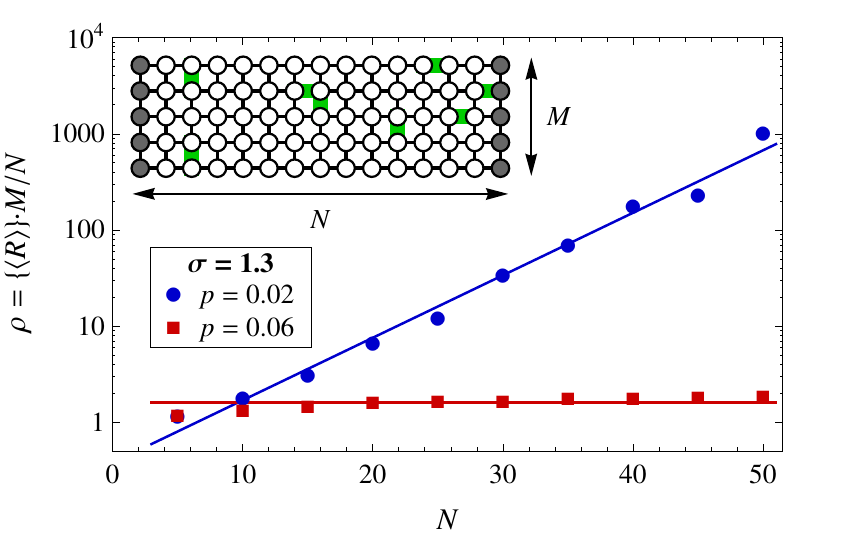}
  \caption{(Color online) Resistivity of a tight-binding ribbon of width $M=5$ and variable length $N$. The
    decoherence induced transition can also be observed in this two-dimensional system. The solid lines are
    fits with an exponential function and a constant, respectively. Numerical averages were taken over $25N
    \cdot 10^5$ decoherence and disorder configurations.}
  \label{fig:4}
\end{figure}

The decoherence induced transition from localized to Ohmic behavior is not restricted to one-dimensional
chains but is found also in two-dimensional tight-binding ribbons, see \fig{4}. In these ribbons decoherence
is introduced by replacing randomly selected bonds with virtual reservoirs, see the marked bonds in the
inset. However, a subdivision into smaller subsystems is in general not possible and the averages can be
calculated only numerically. Note that for these ribbons the critical degree of decoherence is less compared
to one-dimensional chains. This can be understood, by recalling that localization under the effect of
decoherence is caused by the long coherent subsystems. In a ribbon of length $N$ and width $M$ the number of
bonds is $N\lr{2M-1}-M$ and thus, the probability of having a long section in the ribbon, which does not
contain any decoherence reservoir, decreases by a factor $1/(2M-1)$ compared to one-dimensional chains.

\begin{figure}[t]
  \centering
  \includegraphics[scale=1]{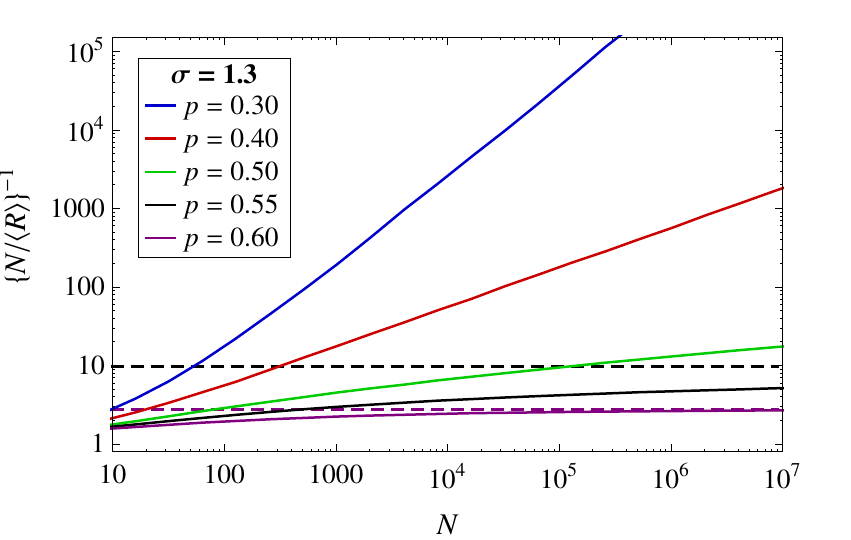}
  \caption{(Color online) Resistivity of a tight-binding chain as a function of its length after averaging
    numerically the conductance over $10^4$ random uncorrelated decoherence configurations. A minimal degree
    of decoherence is necessary for Ohmic conduction, whereas below a power-law divergence is found.  Deep in
    the Ohmic regime, the resistivity is independent of the averaging process, see the convergence of the
    solid curve for $p=0.60$ to the dashed curve, which gives the resistivity due to \eq{15}. However, the
    influence of the averaging process in the transitional regime is not clear, see the curve for $p=0.55$,
    which should converge to the corresponding dashed horizontal line, and the curve for $p=0.50$, which
    should diverge.}
  \label{fig:5}
\end{figure}

Finally we study, if the transition occurs also, when the decoherence average is not performed over the
resistance $\{ \langle R \rangle \}$ but over the conductance $ \{ 1/ \langle R \rangle \}$, which corresponds
to an experiment, where the bias voltage is fixed and the current through the system is measured. In this
case, analytical calculations are demanding but the numerical average over random uncorrelated decoherence
configurations clearly confirms that a minimal degree of decoherence is necessary for Ohmic transport, see
\fig{5}. Below this threshold a power-law divergence is found in contrast to the exponential increase \eq{16}
in the case of averaging the resistance. Figure~\ref{fig:5} also shows that deep in the Ohmic regime, the
resistivity is independent of the averaging process, see the convergence of the solid curve for $p=0.60$ to
the dashed horizontal line, which gives the analytically known limit value \eq{15} in the case of averaging
the resistance. However, from the numerical data it is not clear, whether this is true also in the
transitional regime and whether the critical degree of decoherence depends on the averaging process, see the
curve for $p=0.55$, which should converge to the corresponding dashed horizontal line representing \eq{15},
whereas the curve for $p=0.50$ should diverge. Anyway, for the purpose of this paper it is more important that
the discussed metal-insulator transition appears independently of the averaging process.

\section{Conclusions} \label{sec:Concl}

In this paper, we have shown that the spatial distribution of decoherence, caused by local interactions of the
conduction electrons with their environment, has a significant influence on the transport through a disordered
quantum system. When the decoherence is homogenously distributed, Ohmic conduction is found for any finite
degree of decoherence. In contrast, for random uncorrelated decoherence, a minimal degree of decoherence is
necessary, whereas below the system is localized. This transition is caused by the interplay of the
exponentially increasing coherent resistance \eq{8} and its exponentially decreasing importance \eq{12}. The
other characteristics of our model (one-dimensional chains, complete phase randomization, decoherence average
of the resistance) are not important for this transition, but help to gain much insight into the transport
problem by analytical calculability. To summarize in other words, we have shown that an Anderson insulator can
be stable against decoherence effects, if these are randomly distributed. This result could help to gain
information from resistance measurements on the distribution of the decoherence in a nanosystem.

\begin{acknowledgement}
  This work was supported by Deutsche Forschungsgemeinschaft under Grant No. GRK1240 and
  SPP1386. O.U. acknowledges financial support of the Hungarian NKTH-OTKA Grant No. CNK80991. T.S. thanks
  T. H. Seligman for inspiring discussions and his hospitality at the Centro Internacional de Ciencias A.C. in
  Cuernavaca, M\'exico. We are also grateful to L. Brendel and M. Zilly for useful discussions and helpful
  remarks.
\end{acknowledgement}

\appendix
\section{Disorder averaged resistance of coherent tight-binding chains} \label{sec:DisAvRes}

In this Appendix we calculate analytically the disorder averaged resistance of coherent tight-binding chains
by means of generating functions. Using a recursive scattering approach, Stone et al. \cite{Stone1981} arrived
at the same result.

We consider a chain of length $N$ described by the Hamiltonian \eq{7}, which is connected to two reservoirs by
the self-energy
\begin{equation}
  \label{eq:19}
  \Gm= \nu +\I \eta
\end{equation}
acting on the first and last site of the chain. Because of the tridiagonal structure of $G^{-1}$ the
resistance, defined as the inverse of the transmission \eq{5}, can be calculated recursively in the same way
as in \cite{Zilly2012, Stegmann2012}
\begin{equation}
  \label{eq:20}
  \frac{1}{T_N}= \frac{\Abs{r_N-\Gm r_{N-1} -\Gm s_{N} +\Gm^2 s_{N-1}}^2}{4 \eta^2}
\end{equation}
with the polynomials
\begin{align}
  \label{eq:21}
  & r_{i}=\lr{E-\veps_i}r_{i-1} -r_{i-2}, \quad && s_i=\lr{E-\veps_i}s_{i-1}-s_{i-2}, \notag \\
  & r_{0}=1, \quad && s_1=1, \notag \\
  & r_{-1}=0, \quad && s_0=0.
\end{align}
Using these recursion relations, the disorder averaged resistance can also be calculated recursively
\begin{align}
  \label{eq:22}
  \left\langle \frac{1}{T_N} \right\rangle &= \int \frac{1}{T_N} \prod_{i=1}^N w(\veps_i) \d \veps_i  \notag\\
  &\begin{aligned}
    = \frac{1}{4\eta^2} \Bigl[ R_N  &+ 2\Abs{\Gm}^2 R_{N-1}+\Abs{\Gm}^4 R_{N-2} -4 \nu S_{N} \\[-0.5ex]
    & -4\nu\Abs{\Gm}^2S_{N-1}+4\nu^2 U_N+2\Abs{\Gm}^2 \Bigr],
  \end{aligned}
\end{align}
with
\begin{align}
  \label{eq:23}
  R_N&= \left \langle r_N^2\right \rangle= \int r_N^2 \prod_{i=1}^N w(\veps_i) \d \veps_i \notag\\
  &= \lr{E^2+ \sg^2} R_{N-1} -2E S_{N-1} +R_{N-2},
\end{align}
\begin{align}
  \label{eq:24}
  S_N&= \left \langle r_N r_{N-1}\right \rangle= \int r_N r_{N-1} \prod_{i=1}^N w(\veps_i) \d \veps_i \notag\\
  &=ER_{N-1} -S_{N-1},
\end{align}
\begin{align}
  \label{eq:25}
  U_N&= \left \langle r_N s_{N-1}\right \rangle= \int r_N s_{N-1} \prod_{i=1}^N w(\veps_i) \d \veps_i \notag\\
  &=E S_{N-1} -U_{N-1} -1
\end{align}
and the initial conditions $R_0=1, R_{-1}=S_0=U_1=0$.

In order to solve this recursion, we calculate the generating functions $F_P(z)= \sum_{N=1}^{\infty} P_N
z^{N-1}$ of the polynomials $P \in \{ R,S,U\}$ and with these the generating function
\begin{align}
  \label{eq:26}
  F_{\braket{\frac{1}{T_N}}}(z)&= \sum_{N=1}^{\infty} \left\langle \frac{1}{T_N} \right\rangle z^{N-1} \notag\\[0.5ex]
  &\begin{aligned}
    = \: &\frac{1}{\eta^2} \biggl[ F_R(z) +\bigl( 2\Abs{\Gm}^2 +z\Abs{\Gm}^4 \bigr) \bigl( 1+zF_R(z) \bigr) \\[-0.5ex]
    &-4 \nu\bigl( 1+z\Abs{\Gm}^2 \bigr) F_S(z)+4 \nu^2 F_U(z) +\frac{2\Abs{\Gm}^2}{1-z} \biggr] \notag
  \end{aligned}\\
  &= \frac{1}{2\lr{1-z}} -\frac{1}{4\eta^2}\frac{A(z)}{N_1(z)},
\end{align}
where
\begin{align}
  \label{eq:27}
  A(z)= \Bigl[&1 -2\nu^2 +\Abs{\Gm}^4 \Bigr] z^2 +\Bigl[1 +\bigl( 2\nu^2 -1\bigr) \bigl( E^2-\sg^2 \bigr) \notag\\[-0.5ex]
  &+2\Abs{\Gm}^2 \lr{1-2\nu E} +\Abs{\Gm}^4 \Bigr] z \notag\\[-0.5ex]
  &+E^2+\sg^2 +2\Abs{\Gm}^2-4\nu E+2\nu^2,
\end{align}
\begin{align}
  \label{eq:28}
  N_1(z)&= z^3 -\lr{E^2-\sg^2-1}z^2 +\lr{E^2+\sg^2-1}z-1,
\end{align}
generalizing \cite{Zilly2012} beyond the wide-band approximation. We perform a partial fraction decomposition
of $F_{\braket{1/T_N}}$, or for simplicity rather of
\begin{align}
  \label{eq:29}
  \frac{A(z)}{N_1(z)}&= -4\eta^2 \sum_{N=1}^\infty \lr{\left\langle \frac{1}{T_N} \right\rangle -\frac{1}{2}}z^{N-1} \notag\\
  &= \sum_{k=1}^3 \frac{\ap_k}{z-z_k},
\end{align}
where the $z_k$ are the roots of the polynomial $N_1(z)$ for which Vieta's formulas hold
\begin{align}
  \label{eq:30}
  z_1+z_2+z_3&= E^2-\sg^2-1, \notag\\
  z_1z_2 +z_1z_3+z_2z_3&=E^2+\sg^2-1, \notag\\
  z_1 z_2 z_3&=1.
\end{align}
In the same way, the $\ap_k$ are determined as
\begin{align}
  \label{eq:31}
  \ap_k &= \frac{A(z_k)}{3z_k^2 -2z_k\lr{E^2-\sg^2-1}+E^2+\sg^2-1} \notag\\ 
  &= \frac{A(z_k)}{N_1^{'}(z_k)}.
\end{align}
Using the formal power-series 
\begin{equation}
  \label{eq:32}
  \frac{\ap_k}{z-z_k} = -\frac{\ap_k}{z_k} \sum_{N=1}^\infty \lr{\frac{z}{z_k}}^{N-1}
\end{equation}
in \eq{29}, we get finally the analytical formula
\begin{align}
  \label{eq:33}
  \left\langle \frac{1}{T_N} \right\rangle
  &= \frac{1}{2} +\frac{1}{4\eta^2} \sum_{k=1}^3 \frac{\ap_k}{z_k^N} \notag\\
  &= \frac{1}{2} +\frac{1}{4\eta^2} \sum_{k=1}^3 \ap_ke^{-N\log(z_k)}
\end{align}

This is the main result of this Appendix. It gives, together with $z_k$ from \eq{30} and $\ap_k$ from \eq{31},
the disorder averaged resistance of the coherent tight-binding chain of length $N$, which is connected at its
ends to reservoirs by arbitrary self-energies. To our knowledge such a compact analytical formula, namely a
constant plus a sum of three exponential functions, has never been reported before in the literature.

In the case $E=0$, which is mainly discussed in this paper, the roots are given by
\begin{equation}
  \label{eq:34}
  z_{1,2}= -\frac{\sg^2}{2} \pm \sqrt{\frac{\sg^4}{4}+1}, \qquad z_3= -1,
\end{equation}
and \eq{33} simplifies to \eq{8}.

In the following, we discuss possible values of the roots $z_k$ of the polynomial $N_1(z)$, which determine the
behavior of the exponential functions in \eq{33} and thus, the behavior of the resistance. At first, we note
that $N_1(z)$ is independent of the reservoir's self-energy $\Gm$ and thus, also its roots are independent of
the modeling of the reservoirs \cite{Zilly2012}. From $N_1(z=0)=-1$ and $N_1(z=1)=2\sg^2>0$, we learn that $N_1(z)$
has at least one single real root in the interval $]0,1[$, which is denoted by $z_1$ and leads to the
exponential increase of the resistance. More information on the $z_k$ can be gained by the discriminant
\begin{equation}
  \label{eq:35}
  \Dl= \sg^8 -2\sg^4 \lr{E^4+10E^2-2} +E^2\lr{E^2-4}^3.
\end{equation}
For $\Dl<0$, we have the real root $z_1$ and two complex conjugate roots $z_3=z_2^*$. From the third Vieta
formula we learn that $z_2z_3=\Abs{z_2}^2=1/z_1>1$. Therefore, the complex roots cause by their phase an
oscillation, which is exponentially suppressed with the chain length. For $\Dl \geq 0$ all three roots are
real. Again, we learn from the third Vieta formula $z_2z_3=1/z_1>1$. If $z_2, z_3>0$, only one of them can be
less than $1$. However, two roots in the intervall $]0,1[$ contradict to $N_1(0)=-1$ and $N_1(1)=2\sg^2>0$,
which allows only an odd number of roots in this interval. Therefore both, $z_2$ and $z_3$ are larger $1$. If
$z_2,z_3<0$, only one of them can be in the interval $]{-}1,0[$, which contradicts to $N_1(0)=-1$ and
$N_1(-1)= -2E^2<0$ allowing only an even number of roots in this interval. Therefore both, $z_2$ and $z_3$ are
less than $-1$. In both cases their contributions to the resistance are exponentially suppressed.

To summarize, we have only a single real root $z_1$ in the interval $]0,1[$, which dominates the resistance
for $N \to \infty$
\begin{equation}
  \label{eq:36}
  \left\langle \frac{1}{T_N} \right\rangle= \frac{\ap_1}{4\eta^2}e^{N \Abs{\log(z_1)}}.
\end{equation}
This equation also clarifies that the decoherence induced transition appears also in the case of arbitrary
energies $E$ and self-energies $\Gm$.

\section{Resistivity of disordered tight-binding chains under the effect of decoherence} \label{sec:DphAvRes}

In this Appendix we calculate analytically the resistivity \eq{14} of infinitely long ($N \to \infty$), disordered
tight-binding chains under the effect of decoherence. Using \eq{8} and \eq{12} we obtain 
\begin{align} \label{eq:37} 
  \rho
  &=\sum_{j=1}^{N} \frac{u_j}{N} \left\langle \frac{1}{T_j}\right\rangle \notag\\[0.5ex]
  &\xrightarrow{N \to \infty} \frac{p^2}{2}
  \sum_{j=1}^{\infty} e^{-(j-1)/\ell} \left[1+\ap_+ e^{j/\xi} +\ap_- e^{-j/\xi} (-1)^j \right] \notag\\[0.25ex]
  &\begin{aligned}
    =\frac{p^2}{2} \Biggl( \sum_{j=0}^{\infty} e^{-j/\ell} &+\ap_+ e^{1/\xi} \sum_{j=0}^{\infty}\left[e^{1/\xi -1/\ell} \right]^j\\
    &-\ap_- e^{-1/\xi} \sum_{j=0}^{\infty}\left[-e^{-1/\xi -1/\ell} \right]^j \Biggr).
  \end{aligned}
\end{align}

While the first and third geometric series converge for any finite degree of decoherence $\ell>0$, the second
geometric series converges only if $\xi>\ell$. In this case by performing the sums we get
\begin{equation} \label{eq:38}
\rho=\frac{p^2}{2}\left [
  \frac{1}{1-e^{-1/\ell}} 
  +\frac{\ap_+ e^{1/\xi}}{1-e^{1/\xi -1/\ell}}
  -\frac{\ap_- e^{{-1}/\xi}}{1+e^{-1/\xi -1/\ell}} \right ].
\end{equation}
Substituting \eq{10} in \eq{9}, we can express $\ap_{\pm}$ as a function of the disorder $\sg$
\begin{equation} \label{eq:39}
  2\ap_\pm=1 \pm \frac{1}{\sqrt{\frac{\sg^4}{4}+1}}.
\end{equation}
Using $e^{-1/\ell}=1-p$ and $e^{\pm1/\xi}=\pm\frac{\sigma^2}{2} +\sqrt{\frac{\sg^4}{4}+1}$, we obtain after
straightforward algebra \eq{15}.

If $\xi < \ell$, from the second series in \eq{37} follows directly \eq{16}.

\bibliography{./Stegmann2014}

\end{document}